# NaCl-Assisted Substrate Dependent 2D Planar Nucleated Growth of MoS$_2$

Aditya Singh[1], Monika Moun[1], Madan Sharma[1], Arabinda Barman[2], Ashok Kumar Kapoor[1] and Rajendra Singh[1,2]

[1]Department of Physics, Indian Institute of Technology Delhi, New Delhi, India

[2]Nanoscale Research Facility, Indian Institute of Technology Delhi, New Delhi, India

Email: Corresponding author: adityasingh27993@gmail.com

**Abstract**

Synthesis of large-scale, uniform, easily transferable, and highly crystalline monolayer (1L) molybdenum disulfide (MoS$_2$) on different substrates is a challenge and could decide its suitability for opto-electronic device applications. Herein, we report a facile NaCl-assisted Chemical Vapor Deposition (CVD) synthesis of high-quality MoS$_2$ on amorphous, crystalline and layered substrates. Optical microscopy and Raman spectroscopy show that sapphire and SiO$_2$/Si are suitable substrates for large 1L-MoS$_2$ flakes growth, while mica is excellent for large-area continuous films. Comparatively lesser full-width-at-half-maximum (FWHM) of predominant *A* exciton peak (which is associated with direct band gap at K point) in photoluminescence spectra of 1L-MoS$_2$ on sapphire suggests its high crystalline quality. However, 1L-MoS$_2$ on other substrates, especially on quartz and bare Si show poor crystalline quality. The study depicts that the NaCl assists in the formation of seeding promoter such as water-soluble layer of Na$_2$S and/or Na$_2$SO$_4$ on the substrate that helps in 2D planar nucleation of MoS$_2$. The formation of such intermediate seeding layers also helps in layer transfer owing to its easy water solubility. The study could be utilized for large-scale synthesis of 1L-MoS$_2$ on different substrates for high-performance optoelectronic devices.

## 1. Introduction

Two-dimensional (2D) materials like transition metal dichalcogenides (TMDs) are attracting tremendous attention because of their extraordinary properties in monolayer (1L) form[1]. Specifically, molybdenum disulfide ($MoS_2$), a fellow of TMDs, has layer dependent tunable bandgap (1.8 eV for 1L and 1.2 eV for bulk)[2], high in-plane mobility (40-480 $cm^2$/V-s)[3,4], and large Seebeck coefficient[5]. This makes 1L-$MoS_2$ suitable contestant for flexible devices[4,6], field-effect transistors[4,7], gas sensors[8], and photodetectors[9,10]. Various techniques have been established to synthesize high-quality layered $MoS_2$ on different substrates so far[11]. Among them, conventional Chemical Vapor Deposition (CVD)[12,13] has been extensively used to grow large-scale high-quality uniform crystalline 1L-$MoS_2$ at high temperatures (700-850°C). Wu *et al.*[14] used such vapor-solid method to grow high optical quality 1L-$MoS_2$ on sapphire, $SiO_2$ and glass substrates at 900°C. However, such high-temperature synthesis process puts a limit on direct growth on flexible substrates such as mica. A substantial reduction in growth temperature without compromising material quality is a challenge, and hence, attempts are being made to achieve comparatively low-temperature synthesis.[2]. In this regard, the effectiveness of different growth promoters/catalyst to reduce the synthesis temperature are studied. For instance, Gong *et al.*[15] reported notably reduced-temperature (500ºC) growth of monolayers of $WS_2$ and $MoS_2$ using tellurium as a growth promoter. Moreover, alkali halides such as NaF have shown their potential to lower down the growth temperature for the synthesis of $MoS_2$ films on glass substrates[16]. The study suggested that the crystalline quality of $MoS_2$ was achieved by the formation of an intermediate solution of $Na_2S_x \cdot MoS_2$. Another metal halide, NaCl, has been used as a substrate to grow $MoS_2$ layer by Barreau *et al.*[17]. It discusses the possible formation of intermediate $Na_2MoS_2$ compound which is of the same structure as $MoS_2$ and facilitate as a suitable template for 2D planar nucleation. Furthermore, Kim *et al.*[18] carried out detailed investigations on alkali metal halide (KI or NaCl) assisted growth of $MoS_2$ and suggested that pre-exposing the substrate to alkali metal halides decreases the nucleation density and enhances the 2D planar nucleation process, which eventually enables the growth of larger $MoS_2$ layers. $MoS_2$ flakes synthesized by this method show high crystallinity, high electron mobility (~100 $cm^2$/V-s) and optoelectronic properties equivalent to conventional CVD grown $MoS_2$ single-crystals. Therefore, it is evident from the literature that the use of alkali metal halides not only help in reduction of growth temperature but also facilitates a template for large-flake $MoS_2$ growth by the formation of intermediate compounds. Though such alkali halides effectively enhance the 2D growth of $MoS_2$ layer on the substrate such as sapphire, but there is a lack of systematic study on different substrates of varying structure (crystalline, amorphous and layered). Moreover, the formation of intermediate compounds is a possible speculation while it lacks the concrete evidence. Therefore, a comparative study on alkali halide assisted growth of $MoS_2$ on different substrates and understanding the formation of intermediate growth promoters are of immense interest to improve the quality of $MoS_2$ flakes, and their potential use in optoelectronics.

In the present work, we focus on NaCl-assisted growth of uniform, highly crystalline 1L-$MoS_2$ films and flakes on amorphous, crystalline, and layered-flexible substrates at same growth conditions. On the basis of Raman and photoluminescence measurements, we study the crystallinity, strain, uniformity, and defects in NaCl-assisted 1L-$MoS_2$ on these substrates. Furthermore, we did Raman and X-ray photoelectron spectroscopy (XPS) measurements to understand the role of NaCl in the growth of $MoS_2$. We also elucidate on the intermediate compound formation using Raman measurements. The study reveals that the intermediate compounds $Na_2S$ and $Na_2SO_4$ act as effective growth promoter for CVD growth of 1L-$MoS_2$ film/flakes.

## 2. Experimental

Here, we focus on NaCl-assisted low-temperature CVD of $MoS_2$ on amorphous, crystalline, and layered substrates. It is worth to mention that as in our previous work[19], initially we used Tellurium (Te) to bring down the growth temperature of $MoS_2$ on different substrates such as $SiO_2$/Si, sapphire, mica, bare Si, and quartz substrates. But we could synthesize a few 1L-$MoS_2$ flakes along with bulk flakes only on $SiO_2$/Si at $650^0$C (see supplementary material, **Fig. S1f**). While on other substrates, 1L/2L-$MoS_2$ could not be traced/grown. The detailed Te-assisted experimental investigations and results are discussed in Ref. 18.

In the present work, we have chosen the same $SiO_2$ (300 nm, thermally oxidized silicon)/Si and fused quartz for amorphous, bare Si (n-type, 001) and sapphire (c-plane, $Al_2O_3$) for crystalline, and layered mica [$KAl_2(Si_3Al)O_{10}(OH)_2$] as substrates. Growth of $MoS_2$ on different substrates was carried out using a single zone CVD (MTI Corporation). The growth of $MoS_2$ was achieved by mixing NaCl powder with molybdenum trioxide ($MoO_3$) powder in one crucible placed in the middle (hottest zone) of CVD furnace at reduced temperature and atmospheric pressure. While sulfur (S) powder in another crucible was placed at a lower temperature zone at 13.5 cm away from $MoO_3$ and NaCl mixture. No pre-growth treatment like the use of molecular sieves or seeded growth was done to avoid degrading the electrical properties of flakes[12]. Cleaning of all substrates was performed by acetone then IPA, DI water and nitrogen gas blow, and placed in face-down manner over quartz crucible.

**Fig. 1a** and **1b** represent typical schematics of single-zone CVD setup and grown $MoS_2$ flakes on the substrate, respectively. So, we have chosen same growth parameters as in our previous work[19] *viz.*, S to $MoO_3$ particles ratio ~30:1 (S = 100 mg, $MoO_3$ = 15 mg), and NaCl = 100 mg. The CVD furnace was operated with the ramp rate of 10°C per minute. To purge the tube, 480 sccm of argon (Ar) gas was introduced into the tube for 10 minutes at 300°C, and after that, a continuous flow of 120 sccm was maintained during growth. Purging of tube at 300°C removes existing dust, moisture, preoccupied precursors and other contaminations. At 540 ±10°C and 640 ± 5°C, S starts melting and vaporizing, respectively. NaCl-assisted growth of $MoS_2$ on all substrates was achieved at 650°C for 20 min, and after the growth of $MoS_2$, the system was left for natural cooling.

Blake *et al.*[20] reported that color contrast of graphene from $SiO_2$/Si could be considered as a quantitative technique for deciding the number of layers. Here Nikon Eclipse LV100 optical microscope (OM) was used to differentiate among 1L-, 2L-, and few-layer $MoS_2$. Later on, these observations were confirmed by Raman and photoluminescence spectroscopy using LabRAM HR Evolution (Horiba Scientific) with 514 nm laser at room temperature (RT) on $MoS_2$ flakes to detect fundamental modes ($E^1_{2g}$ and $A_{1g}$)[1] and their bandgaps, respectively. Surface morphology and layer thickness of NaCl-assisted as-grown $MoS_2$ flakes were measured by atomic force microscopy (AFM) (Agilent 5600 LS). Field emission scanning electron microscopy (FESEM) was done using FESEM-Zeiss microscope in backscattering mode and by ZEISS EVO 50 in secondary electron mode. X-ray photoelectron spectroscopy (XPS) was carried out using monochromatic Al Kα X-ray line (energy 1486.7 eV, probe size ~ 1.75 mm × 2.75 mm). Crystallinity, lattice structure, and high resolution transmission electron microscope (HRTEM) images were taken by JEOL/JEM-F200.

## 3. Results and discussion

### 3.1. NaCl-assisted growth of MoS$_2$

NaCl plays a significant role in CVD synthesis of TMDs by reducing the growth temperature[21,22] and forming a smooth surface over substrate[23] or simplifying the deposition of metal-oxyhalide species over the substrate by reacting with metal oxide[21]. Initially, we performed without NaCl-assisted (w/o-NaCl, hereafter) CVD synthesis of MoS$_2$ on sapphire substrate at ~ 900°C that resulted in 1L-MoS$_2$ flakes having very small-size (~1 μm) few-layer MoS$_2$ on its center (**Fig. S1a,b**) (see supplementary information) owing to secondary nucleation sites[24]. On the other hand, NaCl-assisted (w-NaCl, hereafter) CVD of MoS$_2$ on sapphire substrate resulted in a large area high-quality 1L-MoS$_2$ film (~ 1 cm × 1 cm) and flakes (~ 30-100 μm), as shown in OM image (**Fig. 2a**) and SEM image (**Fig. 2f**), respectively. Comparing **Fig. 2a**, with **Fig. S1a,b**, it is evident that using NaCl with MoO$_3$ powder during growth process enhances the flake quality in addition to a significant reduction in growth temperature by 250°C. Furthermore, we tried w/o-NaCl CVD MoS$_2$ growth at ~700°C on flexible mica substrate but, few crack lines, and few-layer triangular MoS$_2$ flakes (< 1 μm) were observed (**Fig. S1c**). However, at 650°C, w-NaCl CVD growth of MoS$_2$ resulted in continuous large-area 1L-MoS$_2$ film (~ 1.5 cm × 1.5 cm) (**Fig. 2b**) and small-size triangular flakes (~ 10 μm) (**Fig. 2g**) with no secondary nucleation sites at the center of flakes as was seen in w/o-NaCl growth.

In w/o-NaCl growth of $MoS_2$ on $SiO_2$/Si substrate, tiny flakes of 2L-$MoS_2$ were observed over the large 1L-$MoS_2$ (**Fig. S1e**), while in case of w-NaCl CVD, uniform continuous monolayers (~ 0.8 cm × 1 cm) of $MoS_2$ (**Fig. c**) and triangles (~ 20-80 μm), with sharp edges were obtained, as shown in SEM image in **Fig. 2h**. Likewise, w/o-NaCl CVD $MoS_2$ growth was carried out on quartz substrate, a few monolayers but mostly few-layer $MoS_2$ triangles at substrate-crucible contact interface (**Fig. S1d**) were obtained. While similar deposition experiment in the presence of NaCl resulted in tiny 1L-$MoS_2$ along with 2L-$MoS_2$ (~5 μm) near substrate-crucible contact interface, as shown in OM images in **Fig. 2d**. It is worth mentioning that the growth was observed only at quartz substrate-crucible interface unlike other substrates used in this study. The proper explanation of such area-specific growth is still elusive. However, we believe that the growth is enhanced at the interface owing to longer interaction time of chemical species. It is observed here that the 1L-$MoS_2$ on quartz was localized in a very small area. This suggests that quartz is not a suitable substrate for $MoS_2$ growth even in the presence of NaCl.

The NaCl-assisted $MoS_2$ deposition was also carried on bare Si substrate. Two nucleation processes governed the $MoS_2$ deposition on Si. One at the substrate-crucible interface where small size, possibly 1L/2L-$MoS_2$ flakes growth were dominant; while 3D nucleation mechanism dictated at the center of the substrate which resulted in a growth of few-layer $MoS_2$ as shown in **Fig. 2i**. It was challenging to locate 1L/2L-$MoS_2$ flakes in an optical microscope in the Raman/PL system as the contrast in the Si substrate was very poor unlike in $SiO_2$/Si substrate. Therefore, precise Raman and PL measurements on 1L/2L-$MoS_2$ flakes on Si could not be executed. The possible growth of 1L/2L-$MoS_2$ at the substrate-crucible interface was predicted based on high-resolution optical microscopy study.

## 3.2. Topographical images of as-grown MoS$_2$ triangular flakes

In order to confirm the structures of MoS$_2$ flakes and to measure the thickness, atomic force microscopy (AFM) was carried out for all the five samples on different substrates. Topographic images of as-grown MoS$_2$ flakes on sapphire, mica, quartz SiO$_2$/Si, and bare Si, are shown in **Fig. 3a-e**, respectively. It is evident from the topography images that the larger flake was formed on the sapphire substrate, unlike on four other substrates. The height profiles were also measured on MoS$_2$ flakes on different substrates, as shown in the inset of **Fig. 3a-e**. The profiles show that the thickness of the flakes vary from 0.76 nm to 0.86 nm suggesting the formation of 1L-MoS$_2$ on sapphire, mica, quartz and SiO$_2$/Si substrates [see inset **Fig. 3a-d**], whereas the height profile of MoS$_2$ flake on bare Si shows 3.2 nm thickness confirming few-layer MoS$_2$ [inset **Fig. 3e**].

## 3.3. Raman spectroscopy

The formation of 1L, 2L or multilayer MoS$_2$ was confirmed on different substrates using optical and atomic force microscopy. However, these measurements only provide a quantitative idea. Therefore, a qualitative understanding of MoS$_2$ layer growth can be achieved using Raman spectroscopy study of the film/flakes as it is sensitive to the vibrational motion of atoms. Here, Raman measurements on multiple triangular flakes and continuous films of MoS$_2$ on each substrate were performed at RT using 514 nm laser to evaluate statistically significant differences as depicted in **Fig. 4a-e**. MoS$_2$ has two first-order fundamental Raman modes, $E^1_{2g}$ and $A_{1g}$, which corresponds to in-plane and out-of-plane vibrational motion of atoms, respectively[25]. The position of these peaks depends on strain, layer thickness, presence of impurities etc.[26,27].

Raman spectroscopy was utilized to enumerate substrate-induced strain in as-grown MoS$_2$. Therefore, as a reference 1L-MoS$_2$ having zero strain, was transferred on TEM grid, over which measurements were done at five different locations as shown in **Fig. 4a**. The average values of modes are found to be $E^1_{2g}$ ~ 383.8 and $A_{1g}$ ~ 403.7 cm$^{-1}$. Moreover, to understand the effect of seeding layer such as Na$_2$S or Na$_2$SO$_4$ (presence of these intermediate phase formation is discussed later) on the MoS$_2$ layer, Raman spectra were recorded on as-grown MoS$_2$ flakes on different substrates as well as on MoS$_2$ layer transferred to same type of substrates. **Fig. 4b** shows the shift of $A_{1g}$ and $E^1_{2g}$ modes of transferred 1L-MoS$_2$ with respect to as-grown w-NaCl 1L-MoS$_2$. The insignificant shift in peak positions ($\leq$ 0.3 cm$^{-1}$) of two modes indicate that underlying Na$_2$S/Na$_2$SO$_4$ layers do not introduce any structural disorder in as-grown MoS$_2$. **Fig. 4c** represents Raman spectra of 1L-MoS$_2$ on SiO$_2$/Si, sapphire, quartz and mica substrates and of few-layer MoS$_2$ on bare Si in the presence of NaCl. The intensity ratio of $A_{1g}$ and $E^1_{2g}$ modes (i.e. $A_{1g}/E^1_{2g}$) were calculated for all the samples and shown in **Fig. S2** (see supplementary). The greater value of $A_{1g}/E^1_{2g}$ ratio in case of sapphire and mica suggest the highly crystalline nature of 1L-MoS$_2$. It is evident from the graph that MoS$_2$/Si shows poor $A_{1g}/E^1_{2g}$ ratio and higher Raman shift between $A_{1g}$ and $E^1_{2g}$. The observation confirms the formation of the poor crystalline and bulk MoS$_2$ on Si. Such poor quality of MoS$_2$ could be attributed to the lattice mismatch of MoS$_2$ and Si which will be discussed in detail in the following section.

Associating the Raman analysis of w-NaCl CVD 1L-MoS$_2$ with suspended 1L-MoS$_2$, a phonon mode stiffening was observed with blue-shift in both $E^1_{2g}$ and $A_{1g}$ modes in each sample. Unlike Buscema *et al.*[28], we observed that both $E^1_{2g}$ and $A_{1g}$ mode are sensitive to substrate as they have a variation of $\geq$ 0.5 cm$^{-1}$ on different substrates [see **Fig. 4d**]. $E^1_{2g}$ and $A_{1g}$ modes for mica have highest blue-shifting while other substrates have nearly the same. This variation can be linked to the lattice mismatch between film and substrate that originates from compressive strain as in the case of tensed graphene whose G and 2D Raman peaks are red-shifted owing to tensile force[29].

However, lesser deviation in $E^1_{2g}$ and $A_{1g}$ peak position of 1L-MoS$_2$ on sapphire than on other substrates show relatively high crystallinity and uniformity over a large area [**Fig. 4d**], and its smallest FWHM of $A_{1g}$ mode [**Fig. 4e**] show the lowest structural disorder[30].

## 3.4. Photoluminescence

Room temperature PL spectroscopy was carried out on 1L-MoS$_2$ flakes and continuous films on different substrates to explore optical properties (**Fig. 5**), and the obtained spectra were fitted with Voigt function[31] to analyze excitons and trions. The most intense peak corresponds to optical bandgap (neutral A exciton) while the peak at higher energy side than A relates to B exciton, generated from the splitting of valence band due to spin-orbit coupling[27,32]. Considering the same analogy as reported by Nakano *et al.*[33], hexagonal lattice of single-crystal sapphire (001) may favor epitaxial growth of MoS$_2$ having a hexagonal structure. Although, *a*-axis lattice constant of sapphire is significantly larger than that of MoS$_2$ but the 2 × 2 superlattice of sapphire and 3 × 3 superlattice of MoS$_2$ has lattice mismatch of only 0.5 %. Furthermore, mica has a layered structure and inter-layer van Der Waals bonding, that may favour the growth of monolayer MoS$_2$ having hexagonal structure too. Ichimiya *et al.*[34] reported that higher the stress, lesser would be A/B exciton ratio. So, these points clearly explain that 1L-MoS$_2$ grown on sapphire has the highest quality PL followed by 1L-MoS$_2$/mica, while 1L-MoS$_2$ on SiO$_2$/Si has the lowest (**Fig. 5e**). The observation is consistent with the Raman analyses of the samples.

## 3.5. UV-Visible absorption

Optical transparency of sapphire, mica, and quartz gave us an advantage in conducting UV-vis absorption measurements on as-grown large area 1L-MoS$_2$ [see **Fig. 6**]. The observed spectra exhibit two prominent *A* and *B* excitonic absorption bands. These bands are originated from valence band splitting triggered by spin-orbit coupling at K point[32]. *A* and *B* excitonic energy difference values from UV-vis of 1L-MoS$_2$ on substrates match with respective valence band splitting values from PL [see **Fig. 5**]. This observation confirms high optical quality, fine electronic structure, and uniformity of our large area 1L-MoS$_2$[26,35]. Therefore, based on Raman, PL, and UV-vis, we confirm the NaCl-assisted high-quality large sized/area growth of 1L-MoS$_2$ on sapphire and mica substrates as compared to exfoliated samples reported earlier[28].

3.6. **X-ray photoelectron spectroscopy**

So far, we investigated the crystallinity and optical properties of as-grown MoS$_2$ films/flakes on different substrates. However, it is important to investigate the chemical properties of the films/flakes to know the quality of the MoS$_2$ growth. Therefore, XPS was carried out on w-NaCl 1L-MoS$_2$ on SiO$_2$/Si substrate. **Fig. 7a** shows the survey spectrum of MoS$_2$ and **Fig. 7b** shows the binding energy (BE) of C 1*s* peak, considered at 284.8 eV for the calibration purpose. The BE of core level orbitals of Mo$^{4+}$ [**Fig. 7c**], 3d$_{5/2}$ and 3d$_{3/2}$ are centered at 230.0 and 233.1 eV, respectively with FWHM of ~1.7 eV which are consistent with the previous reports[36]. The Mo$^{4+}$ 3$d_{5/2}$ has low intense absorption shoulder positioned at 227.3 eV belongs to oxidation of sulfur with -2, and low-intensity peak at 236.0 eV belongs to Mo$^{6+}$ oxidation state, possibly created from the oxidation of sample near the grain boundary and/or other defects[37]. S 2*p* spectrum exhibits spin-orbit doublets (**Fig. 7d**) 2$p_{3/2}$ and 2$p_{1/2}$, centered at 162.9 and 164.2 eV, respectively having energy separation of 1.3 eV. Na 1*s* core-level spectra of w-NaCl MoS$_2$ positioned at 1072.3 eV (**Fig. 7e**) signifies the presence of Na$_2$S or/and Na$_2$SO$_4$ underneath of MoS$_2$ flakes[38,39]. However, when XPS was recorded on w/o-NaCl MoS$_2$, no sign of Na 1*s* core-level spectra observed [see **Fig. 7f**].

## 3.7. NaCl-asisted MoS$_2$ growth mechanism

To understand the growth mechanism, we carried out a separate experiment on w-NaCl and w/o-NaCl growth of MoS$_2$. First, only S and NaCl powder were taken for growth on SiO$_2$/Si substrate at 650°C. We observed [**Fig. 8a**] Raman peaks at 152, 182, and 218 cm$^{-1}$ that correspond to Na$_2$S[40] whereas peaks at 647, 980, and 1152 cm$^{-1}$ referring to Na$_2$SO$_4$[41,42]. Furthermore, we performed MoO$_3$ treatment of the same sample under the same growth condition. We observed Raman peaks at 150, and 185 cm$^{-1}$ [**Fig. 8b**] which belong to Na$_2$S and peaks at 450, 627, and 927 cm$^{-1}$ correspond to Na$_2$SO$_4$. Whereas high-intensity peaks near 400 cm$^{-1}$ are E$^1_{2g}$ (~380 cm$^{-1}$) and A$_{1g}$ (~405 cm$^{-1}$) peaks originates from MoS$_2$[25].

Since Na$_2$S and Na$_2$SO$_4$ compounds are water-soluble[43,44], we could easily transfer the layer of 1L-MoS$_2$ on a different substrate using same process as described by Zhang *et al.*[45]. **Fig. 8c** depicts the Raman spectrum of 1L-MoS$_2$ on sapphire transferred from the same SiO$_2$/Si substrate as shown in **Fig. 8b**. High-intensity peaks near 400 cm$^{-1}$ belong to 1L-MoS$_2$ (E$^1_{2g}$ ~386 cm$^{-1}$ and A$_{1g}$ ~404 cm$^{-1}$) having the peak separation of ~18 cm$^{-1}$ while shoulder peak of MoS$_2$ at ~417 cm$^{-1}$ belongs to sapphire substrate[46]. Here we observed no sign of Na$_2$S and/or Na$_2$SO$_4$ in Raman spectra of transferred MoS$_2$. So, we assume that Na$_2$S and/or Na$_2$SO$_4$ act as a water-soluble sacrificial layer that suppresses nucleation density and enhances the domain size of MoS$_2$ flakes via 2D planar nucleation. It was observed that alkali halides (NaCl and KCl, etc.) act alike seeding promoters similar to 3,4,9,10-perylene tetracarboxylic acid dianhydride (PTCDA), and perylene-3,4,9,10-tetracarboxylic acid tetrapotassium salt (PTAS), etc. which serve well for suppressing nucleation density and enhancing 2D planar nucleation mediated growth in CVD of TMDs[18].

Furthermore, we did Raman measurements on SiO$_2$/Si substrate from where MoS$_2$ was transferred and observed no sign of either MoS$_2$ or Na$_2$S and/or Na$_2$SO$_4$ (**Fig. 8d**). Peaks at 520 and 955 cm$^{-1}$ correspond to first and second order Si Raman peaks[47]. This indicates that Na$_2$S and/or Na$_2$SO$_4$ were dissolved in DI water during the layer transfer process and they have no affiliation with MoS$_2$ after the transfer on a different substrate.

We observed that NaCl play a significant role in CVD of 1L-MoS$_2$ by providing a water-soluble smooth sacrificial layer of Na$_2$S and/or Na$_2$SO$_4$[23,48]. This water-soluble layer makes the transfer of 1L-MoS$_2$ film a smooth and retains its crystallinity. To confirm the crystallinity, same suspended 1L-MoS$_2$ on TEM grid sample that was used for Raman [**Fig. 4a**] was utilized for TEM measurements, HRTEM image, and selected area electron diffraction (SAED). **Fig. 8e** shows honeycomb arrangements of Mo and S atoms, and SAED pattern reveals hexagonal lattice structure of 1L-MoS$_2$ having lattice constant $d = 0.27$ nm that matches with the previous report[12]. The diffraction spots in the SAED pattern are encircled for clarity. So, we firmly believe that low-cost NaCl could act as a suitable replacement of costly catalysts and growth seeding promoters like PTCDA and PTAS for the synthesis of large-area high-quality monolayers of MoS$_2$ in CVD process. In addition, we observed that the use of NaCl enables high-quality growth of 1L-MoS$_2$ in most of the substrates except bare Si.

## 4. Conclusions

In conclusion, NaCl-assisted growth of uniform, highly crystalline 1L-MoS$_2$ films and flakes were carried out on amorphous, crystalline, and layered substrates. The growth temperature significantly reduced from ~900 to 650°C, and the crystalline quality of film/flakes were enhanced as observed from Raman and PL measurements. Optical microscopy shows 1L-MoS$_2$ flakes grown over SiO$_2$/Si and sapphire substrate are large-size and have sharp edges. Raman/PL show that 1L-MoS$_2$/sapphire has relatively better optical properties and least structural disorder than other substrates. It has shown tiny valence band splitting as compared to other substrates. In contrast to large 1L-MoS$_2$ flake growth on sapphire, a continuous film of 1L-MoS$_2$ was observed over entire mica substrate which was attributed to the layered nature of the later. The growth was discussed based on NaCl-assisted formation of seeding layer of Na$_2$S and/or Na$_2$SO$_4$ which enhances 2D planar nucleation. In addition, the water-solubility of this seeding layer was advantageous as it helps to transfer the 1L-MoS$_2$ layer on the desired substrate. Finally, we believe that NaCl could be considered as a cost-effective growth promoter for CVD growth of high-quality, large-area 1L-MoS$_2$ on any desired substrate at a reduced temperature.

**Acknowledgements**

The authors are grateful to the Nanoscale Research Facility (NRF) at the Indian Institute of Technology Delhi for providing the characterization facilities. The partial financial support from the Grand Challenge Project "MBE growth of 2D-Materials" funded by MHRD and IIT Delhi is gratefully acknowledged.

# References

(1) Wang, Q. H.; Kalantar-Zadeh, K.; Kis, A.; Coleman, J. N.; Strano, M. S. Electronics and Optoelectronics of Two-Dimensional Transition Metal Dichalcogenides. *Nat. Nanotechnol.* **2012**, *7* (11), 699.
(2) Kuc, A.; Zibouche, N.; Heine, T. Influence of Quantum Confinement on the Electronic Structure of the Transition Metal Sulfide TS2. *Phys. Rev. B* **2011**, *83* (24), 245213. https://doi.org/10.1103/PhysRevB.83.245213.
(3) Bao, W.; Cai, X.; Kim, D.; Sridhara, K.; Fuhrer, M. S. High Mobility Ambipolar MoS2 Field-Effect Transistors: Substrate and Dielectric Effects. *Appl. Phys. Lett.* **2013**, *102* (4), 042104. https://doi.org/10.1063/1.4789365.
(4) Shao, P.-Z.; Zhao, H.-M.; Cao, H.-W.; Wang, X.-F.; Pang, Y.; Li, Y.-X.; Deng, N.-Q.; Zhang, J.; Zhang, G.-Y.; Yang, Y.; Zhang, S.; Ren, T.-L. Enhancement of Carrier Mobility in MoS2 Field Effect Transistors by a SiO2 Protective Layer. *Appl. Phys. Lett.* **2016**, *108* (20), 203105. https://doi.org/10.1063/1.4950850.
(5) Buscema, M.; Barkelid, M.; Zwiller, V.; van der Zant, H. S. J.; Steele, G. A.; Castellanos-Gomez, A. Large and Tunable Photothermoelectric Effect in Single-Layer MoS2. *Nano Lett.* **2013**, *13* (2), 358–363. https://doi.org/10.1021/nl303321g.
(6) Liu, K.; Yan, Q.; Chen, M.; Fan, W.; Sun, Y.; Suh, J.; Fu, D.; Lee, S.; Zhou, J.; Tongay, S.; Ji, J.; Neaton, J. B.; Wu, J. Elastic Properties of Chemical-Vapor-Deposited Monolayer MoS2, WS2, and Their Bilayer Heterostructures. *Nano Lett.* **2014**, *14* (9), 5097–5103. https://doi.org/10.1021/nl501793a.
(7) Peelaers, H.; Van de Walle, C. G. Effects of Strain on Band Structure and Effective Masses in MoS2. *Phys. Rev. B* **2012**, *86* (24), 241401. https://doi.org/10.1103/PhysRevB.86.241401.
(8) Shokri, A.; Salami, N. Gas Sensor Based on MoS2 Monolayer. *Sens. Actuators B Chem.* **2016**, *236*, 378–385. https://doi.org/10.1016/j.snb.2016.06.033.
(9) Moun, M.; Kumar, M.; Garg, M.; Pathak, R.; Singh, R. Understanding of MoS2/GaN Heterojunction Diode and Its Photodetection Properties. *Sci. Rep.* **2018**, *8*. https://doi.org/10.1038/s41598-018-30237-8.
(10) Lopez-Sanchez, O.; Lembke, D.; Kayci, M.; Radenovic, A.; Kis, A. Ultrasensitive Photodetectors Based on Monolayer MoS2. *Nat. Nanotechnol.* **2013**, *8* (7), 497–501. https://doi.org/10.1038/nnano.2013.100.
(11) Briggs, N.; Subramanian, S.; Lin, Z.; Li, X.; Zhang, X.; Zhang, K.; Xiao, K.; Geohegan, D.; Wallace, R.; Chen, L.-Q.; Terrones, M.; Ebrahimi, A.; Das, S.; Redwing, J.; Hinkle, C.; Momeni, K.; Duin, A. van; Crespi, V.; Kar, S.; Robinson, J. A. A Roadmap for Electronic Grade 2D Materials. *2D Mater.* **2019**, *6* (2), 022001. https://doi.org/10.1088/2053-1583/aaf836.
(12) Tao, L.; Chen, K.; Chen, Z.; Chen, W.; Gui, X.; Chen, H.; Li, X.; Xu, J.-B. Centimeter-Scale CVD Growth of Highly Crystalline Single-Layer MoS2 Film with Spatial Homogeneity and the Visualization of Grain Boundaries. *ACS Appl. Mater. Interfaces* **2017**, *9* (13), 12073–12081. https://doi.org/10.1021/acsami.7b00420.
(13) Yang, P.; Zou, X.; Zhang, Z.; Hong, M.; Shi, J.; Chen, S.; Shu, J.; Zhao, L.; Jiang, S.; Zhou, X.; Huan, Y.; Xie, C.; Gao, P.; Chen, Q.; Zhang, Q.; Liu, Z.; Zhang, Y. Batch Production of 6-Inch Uniform Monolayer Molybdenum Disulfide Catalyzed by Sodium in Glass. *Nat. Commun.* **2018**, *9* (1), 979. https://doi.org/10.1038/s41467-018-03388-5.
(14) Wu, S.; Huang, C.; Aivazian, G.; Ross, J. S.; Cobden, D. H.; Xu, X. Vapor–Solid Growth of High Optical Quality MoS2 Monolayers with Near-Unity Valley Polarization. *ACS Nano* **2013**, *7* (3), 2768–2772. https://doi.org/10.1021/nn4002038.
(15) Gong, Y.; Lin, Z.; Ye, G.; Shi, G.; Feng, S.; Lei, Y.; Elías, A. L.; Perea-Lopez, N.; Vajtai, R.; Terrones, H.; Liu, Z.; Terrones, M.; Ajayan, P. M. Tellurium-Assisted Low-Temperature Synthesis of MoS2 and WS2 Monolayers. *ACS Nano* **2015**, *9* (12), 11658–11666. https://doi.org/10.1021/acsnano.5b05594.
(16) Barreau, N.; Bernède, J. C. Low-Temperature Preparation of MoS2 Thin Films on Glass Substrate with NaF Additive. *Thin Solid Films* **2002**, *403–404*, 505–509. https://doi.org/10.1016/S0040-6090(01)01546-2.
(17) Barreau, N.; Bernède, J. C.; Pouzet, J.; Guilloux-Viry, M.; Perrin, A. Characteristics of Photoconductive MoS2 Films Grown on NaCl Substrates by a Sequential Process. *Phys. Status Solidi A* **2001**, *187* (2), 427–437.
(18) Kim, H.; Ovchinnikov, D.; Deiana, D.; Unuchek, D.; Kis, A. Suppressing Nucleation in Metal–Organic Chemical Vapor Deposition of MoS$_2$ Monolayers by Alkali Metal Halides. *Nano Lett.* **2017**, *17* (8), 5056–5063. https://doi.org/10.1021/acs.nanolett.7b02311.
(19) Singh, A.; Moun, M.; Singh, R. Effect of Different Precursors on CVD Growth of Molybdenum Disulfide. *J. Alloys Compd.* **2019**, *782*, 772–779. https://doi.org/10.1016/j.jallcom.2018.12.230.
(20) Blake, P.; Hill, E. W.; Castro Neto, A. H.; Novoselov, K. S.; Jiang, D.; Yang, R.; Booth, T. J.; Geim, A. K. Making Graphene Visible. *Appl. Phys. Lett.* **2007**, *91* (6), 063124. https://doi.org/10.1063/1.2768624.

(21) Li, S.; Wang, S.; Tang, D.-M.; Zhao, W.; Xu, H.; Chu, L.; Bando, Y.; Golberg, D.; Eda, G. Halide-Assisted Atmospheric Pressure Growth of Large WSe2 and WS2 Monolayer Crystals. *Appl. Mater. Today* **2015**, *1* (1), 60–66. https://doi.org/10.1016/j.apmt.2015.09.001.

(22) Lee, D. K.; Kim, S.; Oh, S.; Choi, J.-Y.; Lee, J.-L.; Yu, H. K. Water-Soluble Epitaxial NaCl Thin Film for Fabrication of Flexible Devices. *Sci. Rep.* **2017**, *7* (1). https://doi.org/10.1038/s41598-017-09603-5.

(23) Zhou, J.; Qin, J.; Guo, L.; Zhao, N.; Shi, C.; Liu, E.; He, F.; Ma, L.; Li, J.; He, C. Scalable Synthesis of High-Quality Transition Metal Dichalcogenide Nanosheets and Their Application as Sodium-Ion Battery Anodes. *J. Mater. Chem. A* **2016**, *4* (44), 17370–17380. https://doi.org/10.1039/C6TA07425A.

(24) Markov, I.; Stoyanov, S. Mechanisms of Epitaxial Growth. *Contemp. Phys.* **1987**, *28* (3), 267–320. https://doi.org/10.1080/00107518708219073.

(25) Li, H.; Zhang, Q.; Yap, C. C. R.; Tay, B. K.; Edwin, T. H. T.; Olivier, A.; Baillargeat, D. From Bulk to Monolayer MoS2: Evolution of Raman Scattering. *Adv. Funct. Mater.* **2012**, *22* (7), 1385–1390. https://doi.org/10.1002/adfm.201102111.

(26) Dumcenco, D.; Ovchinnikov, D.; Marinov, K.; Lazić, P.; Gibertini, M.; Marzari, N.; Sanchez, O. L.; Kung, Y.-C.; Krasnozhon, D.; Chen, M.-W.; Bertolazzi, S.; Gillet, P.; Fontcuberta i Morral, A.; Radenovic, A.; Kis, A. Large-Area Epitaxial Monolayer MoS2. *ACS Nano* **2015**, *9* (4), 4611–4620. https://doi.org/10.1021/acsnano.5b01281.

(27) Chae, W. H.; Cain, J. D.; Hanson, E. D.; Murthy, A. A.; Dravid, V. P. Substrate-Induced Strain and Charge Doping in CVD-Grown Monolayer MoS$_2$. *Appl. Phys. Lett.* **2017**, *111* (14), 143106. https://doi.org/10.1063/1.4998284.

(28) Buscema, M.; Steele, G. A.; van der Zant, H. S. J.; Castellanos-Gomez, A. The Effect of the Substrate on the Raman and Photoluminescence Emission of Single-Layer MoS2. *Nano Res.* **2014**, *7* (4), 561–571. https://doi.org/10.1007/s12274-014-0424-0.

(29) Ni, Z. H.; Yu, T.; Lu, Y. H.; Wang, Y. Y.; Feng, Y. P.; Shen, Z. X. Uniaxial Strain on Graphene: Raman Spectroscopy Study and Band-Gap Opening. *ACS Nano* **2008**, *2* (11), 2301–2305. https://doi.org/10.1021/nn800459e.

(30) Islam, M. R.; Kang, N.; Bhanu, U.; Paudel, H. P.; Erementchouk, M.; Tetard, L.; Leuenberger, M. N.; Khondaker, S. I. Tuning the Electrical Property via Defect Engineering of Single Layer MoS2 by Oxygen Plasma. *Nanoscale* **2014**, *6* (17), 10033–10039. https://doi.org/10.1039/C4NR02142H.

(31) Kaplan, D.; Gong, Y.; Mills, K.; Swaminathan, V.; Ajayan, P. M.; Shirodkar, S.; Kaxiras, E. Excitation Intensity Dependence of Photoluminescence from Monolayers of MoS2 and WS2/MoS2 Heterostructures. *2D Mater.* **2016**, *3* (1), 015005. https://doi.org/10.1088/2053-1583/3/1/015005.

(32) Splendiani, A.; Sun, L.; Zhang, Y.; Li, T.; Kim, J.; Chim, C.-Y.; Galli, G.; Wang, F. Emerging Photoluminescence in Monolayer MoS2. *Nano Lett.* **2010**, *10* (4), 1271–1275. https://doi.org/10.1021/nl903868w.

(33) Nakano, M.; Wang, Y.; Kashiwabara, Y.; Matsuoka, H.; Iwasa, Y. Layer-by-Layer Epitaxial Growth of Scalable WSe2 on Sapphire by Molecular Beam Epitaxy. *Nano Lett.* **2017**, *17* (9), 5595–5599. https://doi.org/10.1021/acs.nanolett.7b02420.

(34) Ichimiya, M.; Watanabe, M.; Ohata, T.; Hayashi, T.; Ishibashi, A. Effect of Uniaxial Stress on Photoluminescence in GaN and Stimulated Emission in InxGa1−xN/GaN Multiple Quantum Wells. *Phys. Rev. B* **2003**, *68* (3), 035328. https://doi.org/10.1103/PhysRevB.68.035328.

(35) Ji, Q.; Zhang, Y.; Gao, T.; Zhang, Y.; Ma, D.; Liu, M.; Chen, Y.; Qiao, X.; Tan, P.-H.; Kan, M.; Feng, J.; Sun, Q.; Liu, Z. Epitaxial Monolayer MoS2 on Mica with Novel Photoluminescence. *Nano Lett.* **2013**, *13* (8), 3870–3877. https://doi.org/10.1021/nl401938t.

(36) Kim, I. S.; Sangwan, V. K.; Jariwala, D.; Wood, J. D.; Park, S.; Chen, K.-S.; Shi, F.; Ruiz-Zepeda, F.; Ponce, A.; Jose-Yacaman, M.; Dravid, V. P.; Marks, T. J.; Hersam, M. C.; Lauhon, L. J. Influence of Stoichiometry on the Optical and Electrical Properties of Chemical Vapor Deposition Derived MoS2. *ACS Nano* **2014**, *8* (10), 10551–10558. https://doi.org/10.1021/nn503988x.

(37) Kim, H.; Dumcenco, D.; Frégnaux, M.; Benayad, A.; Chen, M.-W.; Kung, Y.-C.; Kis, A.; Renault, O. Free-Standing Electronic Character of Monolayer MoS2 in van Der Waals Epitaxy. *Phys. Rev. B* **2016**, *94* (8), 081401. https://doi.org/10.1103/PhysRevB.94.081401.

(38) Song, J.-G.; Hee Ryu, G.; Kim, Y.; Je Woo, W.; Yong Ko, K.; Kim, Y.; Lee, C.; Oh, I.-K.; Park, J.; Lee, Z.; Kim, H. Catalytic Chemical Vapor Deposition of Large-Area Uniform Two-Dimensional Molybdenum Disulfide Using Sodium Chloride. *Nanotechnology* **2017**, *28* (46), 465103. https://doi.org/10.1088/1361-6528/aa8f15.

(39) Turner, N. H.; Murday, J. S.; Ramaker, D. E. Quantitative Determination of Surface Composition of Sulfur Bearing Anion Mixtures by Auger Electron Spectroscopy. *Anal. Chem.* **1980**, *52* (1), 84–92. https://doi.org/10.1021/ac50051a021.

(40) El Jaroudi, O.; Picquenard, E.; Gobeltz, N.; Demortier, A.; Corset, J. Raman Spectroscopy Study of the Reaction between Sodium Sulfide or Disulfide and Sulfur: Identity of the Species Formed in Solid and Liquid Phases. *Inorg. Chem.* **1999**, *38* (12), 2917–2923. https://doi.org/10.1021/ic9900096.

(41) Prieto-Taboada, N.; Vallejuelo, S. F.-O. de; Veneranda, M.; Lama, E.; Castro, K.; Arana, G.; Larrañaga, A.; Madariaga, J. M. The Raman Spectra of the Na2SO4-K2SO4 System: Applicability to Soluble Salts Studies in Built Heritage. *J. Raman Spectrosc.* **2019**, *50* (2), 175–183. https://doi.org/10.1002/jrs.5550.

(42) Mabrouk, K. B.; Kauffmann, T. H.; Aroui, H.; Fontana, M. D. Raman Study of Cation Effect on Sulfate Vibration Modes in Solid State and in Aqueous Solutions. *J. Raman Spectrosc.* **2013**, *44* (11), 1603–1608. https://doi.org/10.1002/jrs.4374.

(43) PubChem. Sodium sulfide https://pubchem.ncbi.nlm.nih.gov/compound/237873 (accessed Jul 31, 2019).

(44) Bharmoria, P.; Gehlot, P. S.; Gupta, H.; Kumar, A. Temperature-Dependent Solubility Transition of Na2SO4 in Water and the Effect of NaCl Therein: Solution Structures and Salt Water Dynamics. *J. Phys. Chem. B* **2014**, *118* (44), 12734–12742. https://doi.org/10.1021/jp507949h.

(45) Zhang, L.; Wang, C.; Liu, X.-L.; Xu, T.; Long, M.; Liu, E.; Pan, C.; Su, G.; Zeng, J.; Fu, Y.; Wang, Y.; Yan, Z.; Gao, A.; Xu, K.; Tan, P.-H.; Sun, L.; Wang, Z.; Cui, X.; Miao, F. Damage-Free and Rapid Transfer of CVD-Grown Two-Dimensional Transition Metal Dichalcogenides by Dissolving Sacrificial Water-Soluble Layers. *Nanoscale* **2017**, *9* (48), 19124–19130. https://doi.org/10.1039/C7NR06928F.

(46) Kadleíková, M.; Breza, J.; Veselý, M. Raman Spectra of Synthetic Sapphire. *Microelectron. J.* **2001**, *32* (12), 955–958. https://doi.org/10.1016/S0026-2692(01)00087-8.

(47) Parker, J. H.; Feldman, D. W.; Ashkin, M. Raman Scattering by Silicon and Germanium. *Phys. Rev.* **1967**, *155* (3), 712–714. https://doi.org/10.1103/PhysRev.155.712.

(48) Wang, Z.; Xie, Y.; Wang, H.; Wu, R.; Nan, T.; Zhan, Y.; Sun, J.; Jiang, T.; Zhao, Y.; Lei, Y.; Yang, M.; Wang, W.; Zhu, Q.; Ma, X.; Hao, Y. NaCl-Assisted One-Step Growth of $MoS_2$–$WS_2$ in-Plane Heterostructures. *Nanotechnology* **2017**, *28* (32), 325602. https://doi.org/10.1088/1361-6528/aa6f01.

# **Figures**

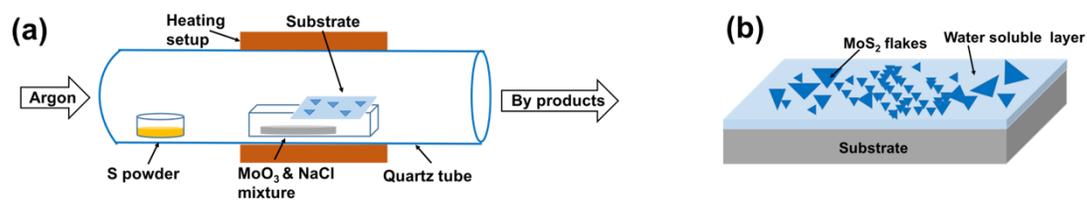

**Figure 1.** Schematic of (a) typical single-zone CVD setup and (b) grown MoS$_2$ flakes on the water-soluble sacrificial layer of Na$_2$S/Na$_2$SO$_4$ on the substrate.

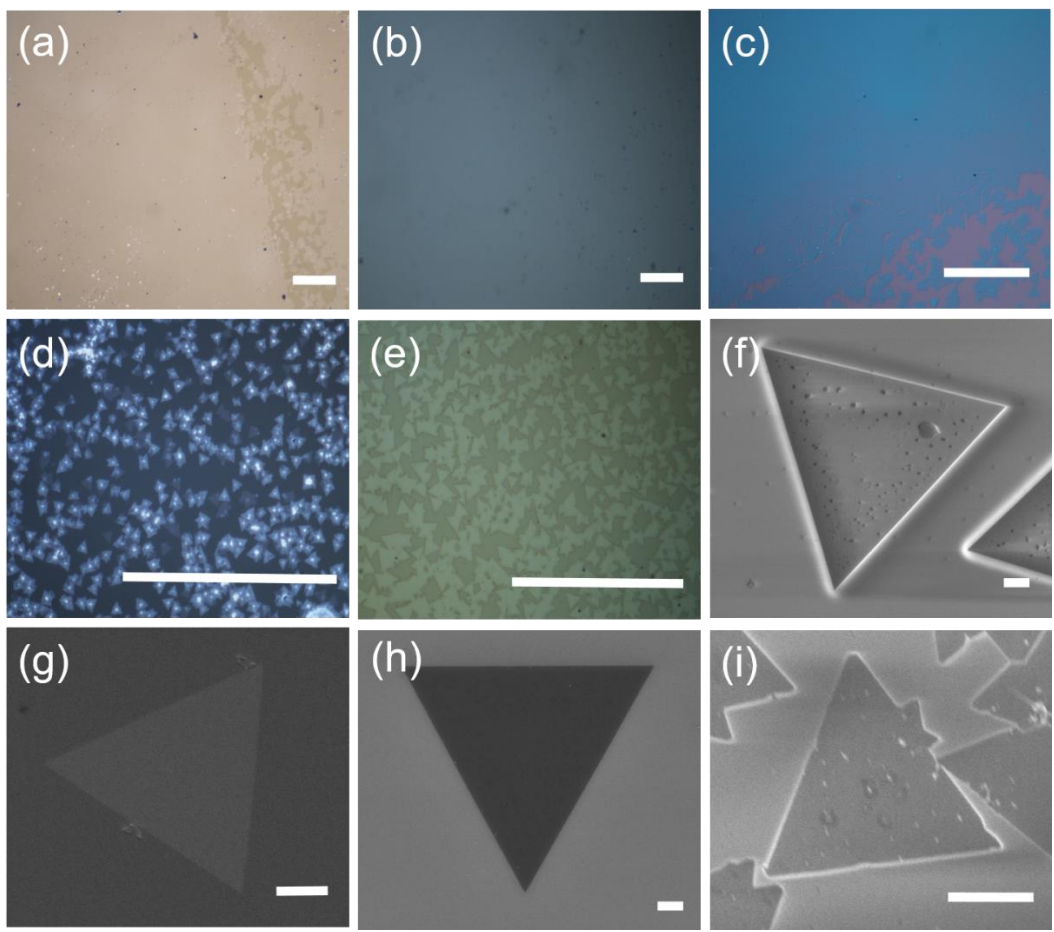

**Figure 2.** OM image of (a)-(c) continuous film of 1L-MoS$_2$ on sapphire, mica and SiO$_2$/Si substrate, respectively and (d), (e) 1L- and few-layer MoS$_2$ on fused quartz and bare silicon, respectively. (f)-(i) SEM images of 1L-MoS$_2$ triangle on sapphire, mica, SiO$_2$/Si, and quartz. Scale bar (a)-(e) is 100 μm and (f)-(i) 2 μm.

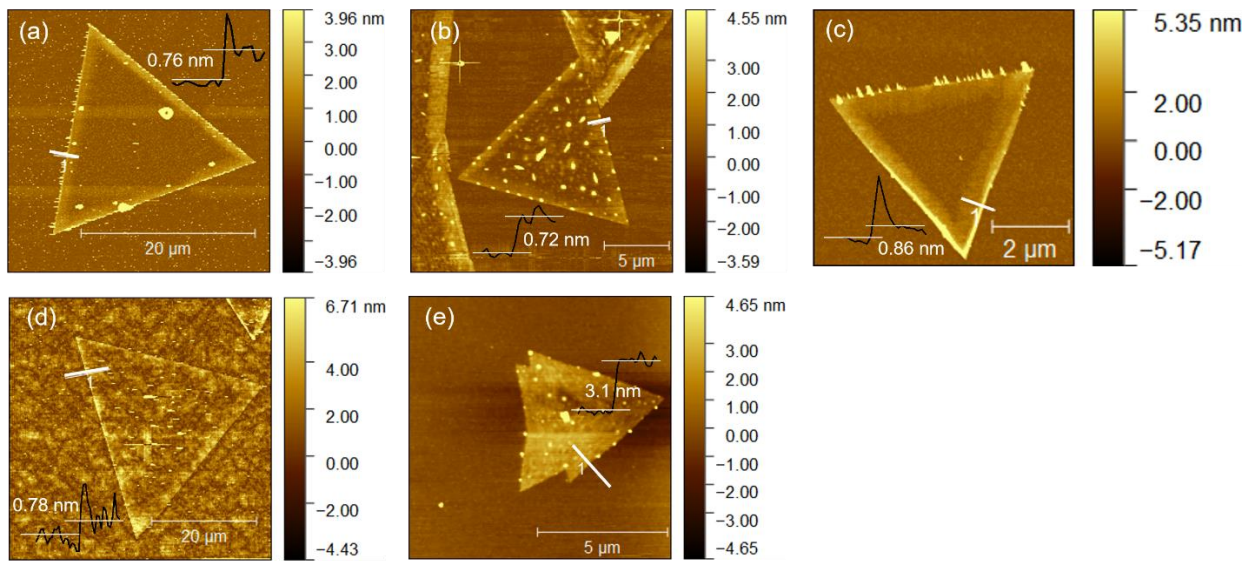

**Figure 3.** AFM images of 1L- MoS$_2$ on (a) sapphire, (b) mica (c) quartz (d) SiO$_2$/Si and few-layer MoS$_2$ on (e) bare Si substrates.

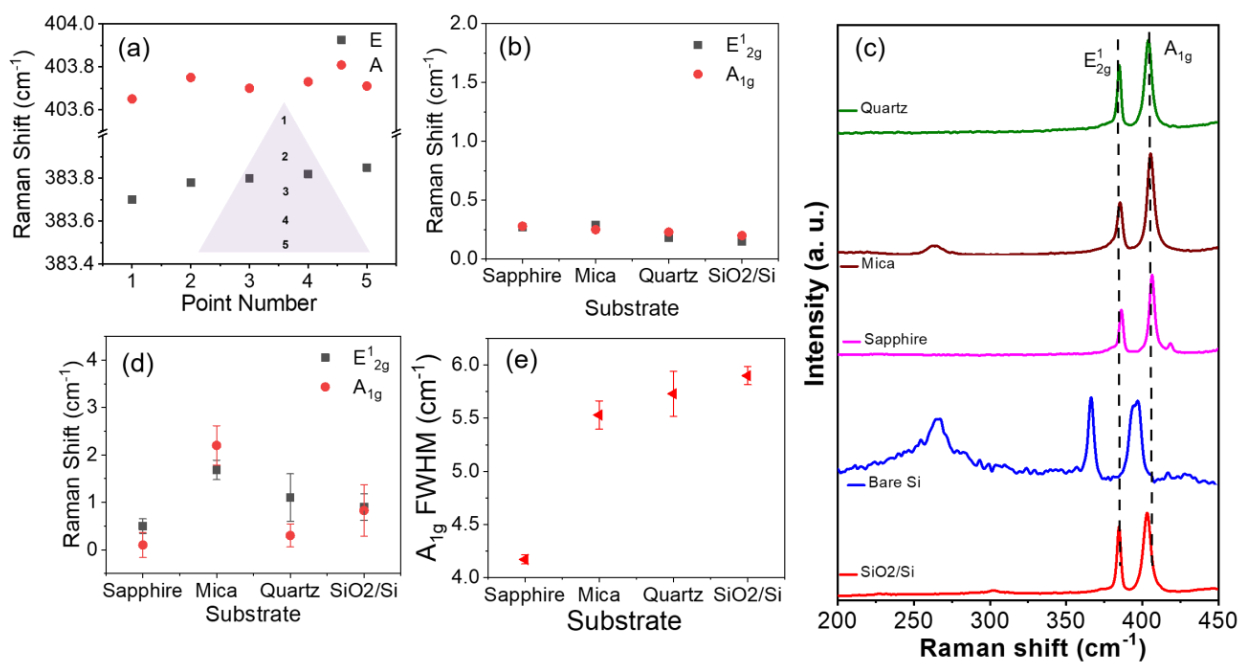

**Figure 4.** Raman spectroscopy analysis of w-NaCl CVD 1L-MoS$_2$ grown on different substrates: (a) Position of A$_{1g}$ and E$^1_{2g}$ Raman modes at various points on suspended CVD 1L-MoS$_2$ on TEM grid and inserted schematic shows points (1-5) where corresponding measurements were done. (b) Shift of A$_{1g}$ and E$^1_{2g}$ modes after transfer to same type substrate with respect to as-grown w-NaCl CVD 1L-MoS$_2$. (c) Raman spectrum of 1L-MoS$_2$ grown on SiO$_2$/Si, sapphire, mica, quartz and few-layer MoS$_2$ on bare Si. (d) Shift of A$_{1g}$ and E$^1_{2g}$ modes of w-NaCl CVD 1L-MoS$_2$ from suspended 1L-MoS$_2$ (E$^1_{2g}$ ~ 383.8 cm$^{-1}$ and A$_{1g}$ ~ 403.7 cm$^{-1}$) and (e) FWHM of A$_{1g}$ Raman mode for different substrates.

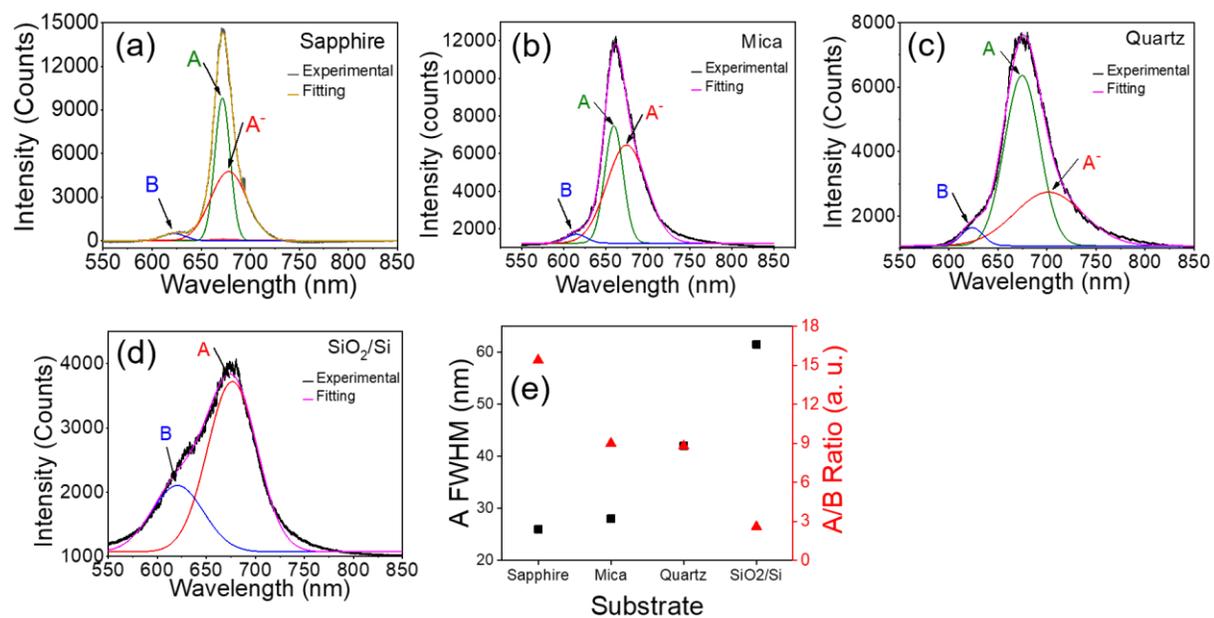

**Figure 5.** Photoluminescence (PL) spectra of 1L-MoS$_2$ on different substrates: 1L-MoS$_2$ on (a) sapphire, (b) mica, (c) quartz, and (d) SiO$_2$/Si substrate. (e) Variation of FWHM of A exciton and ratio of A and B exciton with substrates.

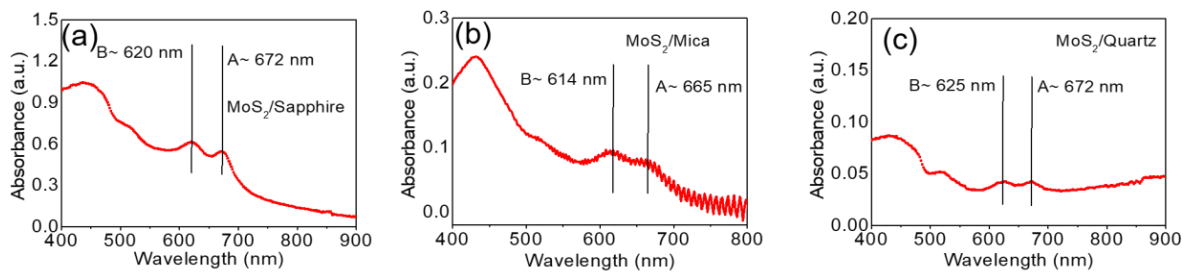

**Figure 6.** UV-vis spectra of 1L-MoS$_2$ on (a) sapphire (b) mica, and (c) quartz substrate.

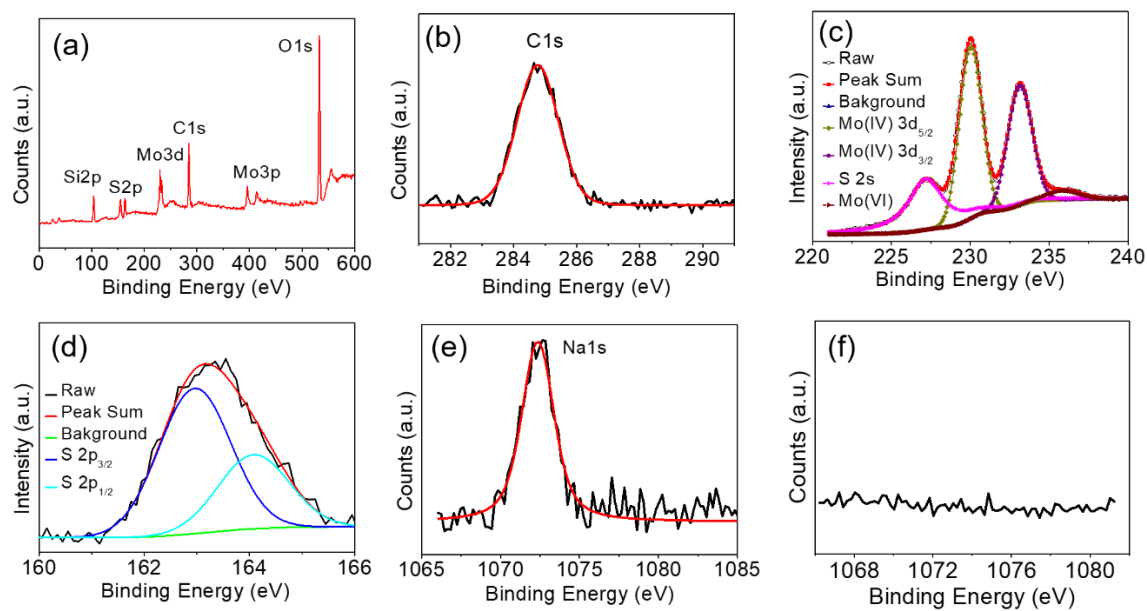

**Figure 7.** XPS of directly grown 1L-MoS$_2$ on SiO$_2$/Si substrate: (a) survey spectrum, (b) core level C 1s peak, (c) Mo 3d peaks for the orbitals 3d$_{5/2}$ and 3d$_{3/2}$ are centered at 230.0 and 233.1 eV, respectively. Peak positioned at 227.2 eV belongs to oxidation of sulfur with -2, and low-intensity peak at 235.9 eV belongs to Mo$^{6+}$ state. (d) S 2p peaks for the orbitals 2p$_{3/2}$ and 2p$_{1/2}$ are centered at 162.9 and 164.2 eV, respectively. (e) Na 1s core-level spectra of NaCl-assisted growth of MoS$_2$ signifies the presence of Na$_2$S or/and Na$_2$SO$_4$ underneath of MoS$_2$ flakes. (f) No Na 1s core-level spectra observed in transferred NaCl-assisted grown MoS$_2$.

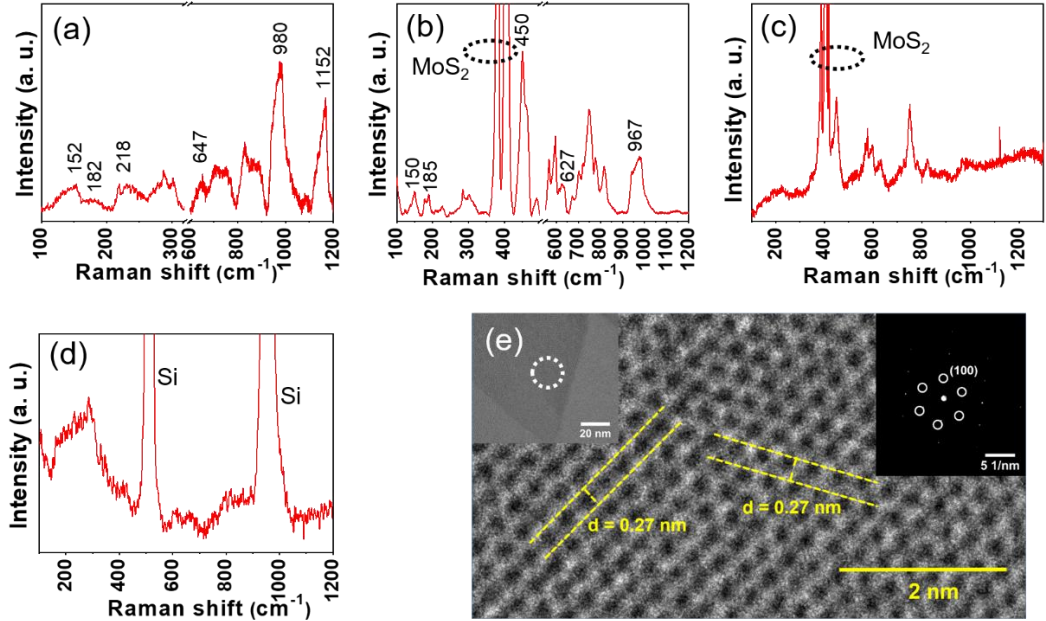

**Figure 8.** (a) Raman spectrum of the as-grown film when only S and NaCl were used. Raman peaks at 152, 182, and 218 cm$^{-1}$ correspond to Na$_2$S whereas peaks at 647, 980, and 1152 cm$^{-1}$ correspond to Na$_2$SO$_4$. (b) Raman spectrum of S and NaCl treated sample exposed under MoO$_3$ powder, shows sign of MoS$_2$, Na$_2$S, and Na$_2$SO$_4$. (c) Raman spectrum of MoS$_2$/sapphire transferred from SiO$_2$/Si substrate shows no sign of Na$_2$S and Na$_2$SO$_4$. (d) Raman spectrum of SiO$_2$/Si substrate from where MoS$_2$ film was transferred. Peaks at 520 and 955 cm$^{-1}$ correspond to first and second order Si Raman peaks. (e) HRTEM image of 1L-MoS$_2$ acquired at circled location in TEM image (top-left) and corresponding SAED pattern (top-right).

**NaCl-Assisted Substrate Dependent 2D Planar Nucleated Growth of MoS$_2$**

Aditya Singh[1], Monika Moun[1], Madan Sharma[1], Arabinda Barman[2], Ashok Kumar Kapoor[1] and Rajendra Singh[1,2]

[1]Department of Physics, Indian Institute of Technology Delhi, New Delhi, India

[2]Nanoscale Research Facility, Indian Institute of Technology Delhi, New Delhi, India

Email: Corresponding author: adityasingh27993@gmail.com

Initially, for low-temperature growth, we did Te-assisted growth and found that 1L-MoS$_2$ flakes were obtained along with few-layer MoS$_2$, this leads to non-uniformity of growth [Fig. S1(f)].

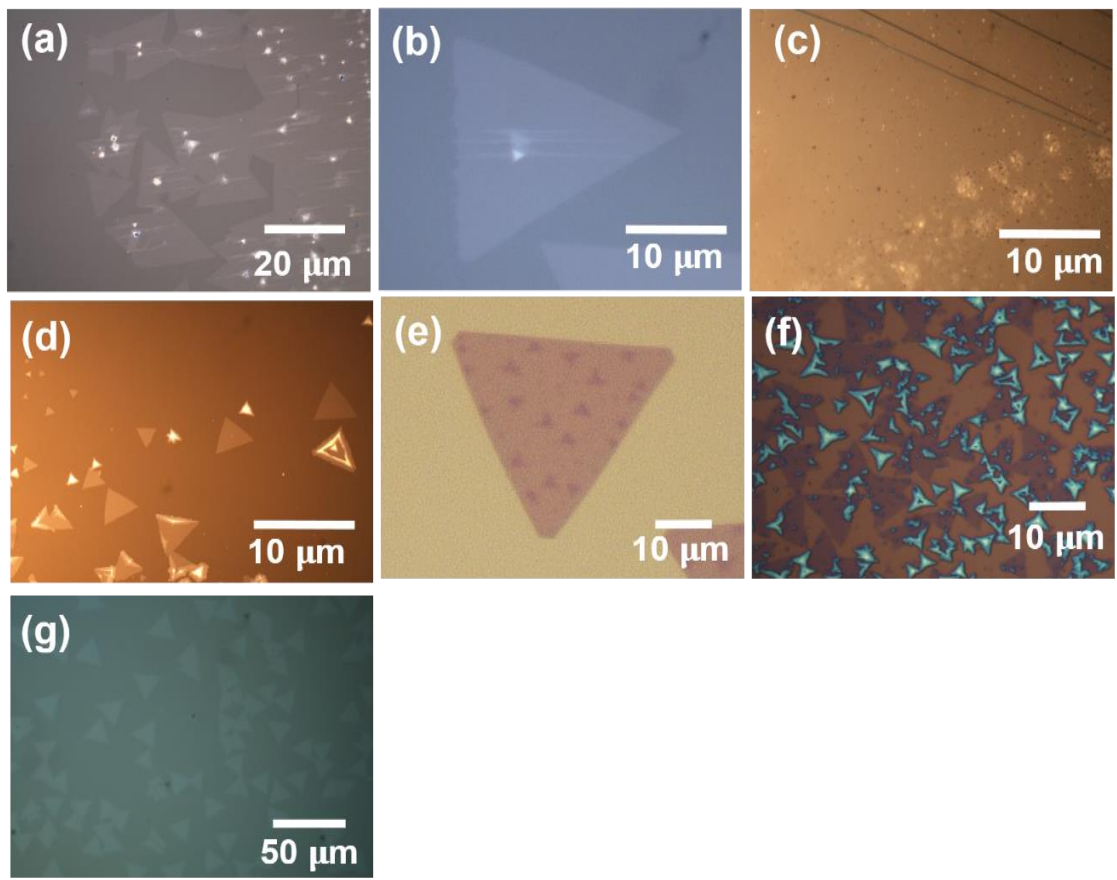

**Figure S1: Optical microscopy (OM) images of MoS$_2$:** w/o-NaCl CVD MoS$_2$ on (a), (b) sapphire, (c) mica, (d) quartz, and (e) SiO$_2$/Si substrate. (f) optical image of Te-assisted CVD (650°C) of MoS$_2$ on SiO$_2$/Si substrate showing mono- and multi-layer MoS$_2$. (g) optical image of 1L or 2L-MoS$_2$ on bare Si substrate.

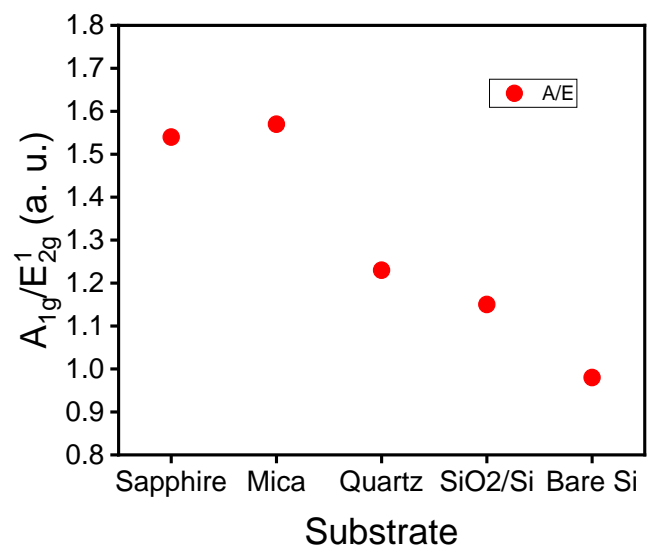

**Figure S2:** Ratio of Intensity of $A_{1g}$ and $E^1_{2g}$ Raman Modes for different substrates.